\documentclass[11pt,preprint]{aastex}
\usepackage{natbib}
\title{The Swift BAT Survey Detects Two Optical Broad Line, X-ray Heavily Obscured Active Galaxies: NVSS 193013+341047 and IRAS 05218-1212}
\author{J. Drew Hogg\altaffilmark{1}, Lisa M. Winter\altaffilmark{1,3},  Richard F. Mushotzky\altaffilmark{2}, \\ Christopher S. Reynolds\altaffilmark{2},  Margaret Trippe\altaffilmark{2}}
\email{james.hogg@colorado.edu, lisa.winter@colorado.edu}

\altaffiltext{1}{Department of Astrophysical \& Planetary Sciences, University of Colorado, UCB 391, Boulder, CO 80309, USA}
\altaffiltext{2}{Department of Astronomy, University of Maryland, College Park, MD 20742, USA}
\altaffiltext{3}{Hubble Fellow}

\begin{abstract}
The Swift Burst Alert Telescope (BAT) is discovering interesting new objects while monitoring the sky in the 14--195\,keV band.  Here we present the X-ray properties and spectral energy distributions for two unusual AGN sources.  Both  NVSS 193013+341047 and IRAS 05218-1212 are absorbed,  Compton-thin, but heavily obscured (N$_{\rm H} \sim 10^{23}$\,cm$^{-2}$), X-ray sources at redshifts $< 0.1$.  
The spectral energy distributions reveal these galaxies to be very red, with high extinction in the optical and UV.  A similar SED is seen for the extremely red objects (EROs) detected in the higher redshift universe.  This suggests that these unusual BAT-detected sources are a low-redshift ($z << 1$) analog to EROs, which recent evidence suggests are a class of the elusive type II quasars.  Studying the multi-wavelength properties of these sources may reveal the properties of their high redshift counterparts.
\end{abstract}

\begin{document}

\section{Introduction}
Super massive black holes are believed to exist at the centers of most massive galaxies. 
While a connection between the central black hole and larger host galaxy has been observationally proven (the M -- $\sigma$ relation), there are still many unanswered questions as to how the central black hole affects the host galaxy's formation and evolution.  Since previous studies of active galactic nuclei (AGN) have relied heavily on optically or X-ray selected samples, they missed a significant fraction ($\approx 20$\%) of AGNs -- the heavily obscured sources 
\citep{2007ApJ...664L..79U,2008ApJ...674..686W,2009ApJ...690.1322W}.  Thus, our understanding of the AGNs and their connection to the host is missing a potentially key parameter.

Surveys of active galactic nuclei are
typically dominated by two selection effects: dilution by starlight
from the host galaxy and obscuration by dust and gas in the host
galaxy and the AGN itself (see \citealt{1994PASP..106..113H} and \citealt{2004ASSL..308...53M}).  These factors previously kept an unbiased AGN sample from
reach.  However, with the capabilities of Swift's Burst Alert
Telescope (BAT), this has changed. The BAT surveys the sky in the hard
X-ray range of 14 -- 195\,keV.  In the first 22 months of the survey, BAT detected $\sim$250
AGN with a median redshift of 0.03 \citep{2010ApJS..186..378T}.  These AGN
were selected purely by their hard X-ray flux, and thus, are not
affected by obscuration by gas and dust below $\approx 10^{24}$\,cm$^{-2}$, which prevents them from
being easily detected in optical or soft band X-ray surveys.

In the BAT sample there is no bias against detecting heavily
obscured but still Compton thin AGNs, thus the sample offers a unique test of the unified model of
AGNs \citep{1993ARA&A..31..473A}.  In the unified model, differences between broad (type 1) and
narrow-line (type 2) AGNs are attributed to differences in the viewing angle to the
central, obscured region \citep{1978PNAS...75..540O}. Therefore, there should be
no differences between the X-ray and optical AGN classifications.

However, the X-ray/optical properties of two sources out of the $\approx 250$ detected do not fit the most basic prediction of the unified model.  Both NVSS 193013+341047 and IRAS 05218-1212 show broad optical lines indicative of a type 1 -- 1.5 optical source, but the low resolution Swift XRT spectra indicated that the sources were likely highly absorbed.  Further, the combined BAT and XRT X-ray spectra indicated that the sources were mildly ``Compton-thick'', with column densities of N$_{\rm H} > 1.4 \times 10^{24}$\,cm$^{-2}$.  The spectral energy distributions (SEDs) of these sources are extremely red, showing very heavy extinction levels in the UV and soft X-rays ($< 2$\,keV).  Such SEDs have only previously been seen for the high redshift ($z \ga 1$) extremely red objects (EROs).  In order to determine the X-ray properties of these sources, we obtained X-ray follow-ups with XMM-Newton.  In this paper, we present both the X-ray and broad-band properties of these two unusual sources detected in the Swift BAT survey.

\section{X-ray Spectra}
\subsection{Observation Details}
XMM-Newton follow-ups were acquired in the 0.1-10 keV band for both NVSS 193013+341047 and IRAS 05218-1212.  Basic details of the sources and the XMM-Newton observations are included in Table~\ref{tbl-1}.  The spectra were processed and extracted using the Science Analysis System (SAS) version 9.0.  We followed the process for reduction outlined in the XMM-Newton ABC Guide\footnote{The ABC Guide is available online at \url{http://heasarc.nasa.gov/docs/xmm/abc/}}.  First, we created calibrated photon event files for the EPIC MOS and pn cameras using the observation data files (ODFs) with the commands {\tt emchain} and {\tt epchain}, respectively. 

The events tables were filtered using standard filters as outlined in the ABC Guide.  For MOS data, good events constitute those with a pulse height in the range of 0.2-12 keV and event patterns that are characterized as 0-12 (single, double, triple, and quadruple pixel events).  For the pn camera, only patterns of 0-4 (single and double pixel events) are kept, with the energy range for the pulse height set between 0.2 and 15 keV.  Bad pixels and events too close to the edges of the CCD chips were rejected using the most stringent selection expression, ÒFLAG == 0Ó, for both data sets.  Based on an inspection of the light curve, we found no evidence of flaring in either observation.  Therefore, we did not apply any temporal filtering.

	The spectra of the sources were extracted using the SAS task {\tt evselect}.   The spectra were extracted in a circular aperture centered on the source.  Additionally, background spectra were extracted with a circular aperture of the same size, in a region nearby the source on the same CCD chip, but free of additional X-ray sources.  Response files were generated using the SAS tasks {\tt rmfgen} and {\tt arfgen}.  The spectra were then binned by 20 counts per bin using {\tt grppha}.

	In addition to the XMM-Newton spectra, we also obtained Swift BAT spectra.  The spectra are time-averaged over 22-months and are fully described in \citet{2010ApJS..186..378T}\footnote{Swift BAT spectra for the 22-month sample are publicly available at \url{http://heasarc.nasa.gov/docs/swift/results/bs22mon/}.}.  The BAT spectra cover the 14--195\,keV band and consist of eight energy bins with edges of 14, 20, 24, 35, 50, 75, 100, and 150 keV.  
	


\subsection{XMM-Newton Light Curve Analysis}

	In order to search for variability in each of the sources, we created background subtracted light curves for pn in a soft (0.3--3\,keV), medium (3--7\,keV), and hard (7--10\,keV) band, using a bin size of 128\,s.  The same source and background regions were used as those to extract the spectra.  For NVSS 193013+341047 the average background subtracted count rate in each band was $1.88 \times 10^{-2}$ cts\,s$^{-1}$ (soft), $4.79 \times 10^{-2}$\,cts\,s$^{-1}$ (medium), and $1.69 \times 10^{-2}$\,cts\,s$^{-1}$ (hard).   For IRAS 05218-1212, the average count rate was $5.15 \times 10^{-2}$\,cts\,s$^{-1}$ (soft), $4.67 \times 10^{-2}$\,cts\,s$^{-1}$ (medium), and $9.98 \times 10^{-3}$\,cts\,s$^{-1}$ (hard).  
	
As a simple search for variability, we computed $\chi^2 = \sum(C_i^2 - \mu^2)/\sigma^2$ for each light curve, where $C_i$ is the background subtracted count rate in each bin, $\mu$ is the average count rate, and $\sigma$ is the error.  For NVSS 193013+341047, the reduced $\chi^2$ values are 0.85 (soft), 0.69 (medium), and 1.13 (hard), where the light curves had 113 dof.  For IRAS 05218-1212, the reduced $\chi^2$ values are 0.77 (soft), 1.09 (medium), and 1.09 (hard), where the light curves had 101 dof.  The amount of variability is low in all of the light curves.  However, the medium band, which encompasses the Fe K$\alpha$ emission region, corresponds to the greatest variability in NVSS 193013+341047 and the soft band corresponds to the greatest variability in IRAS 05218-1212.  The band with the greatest short-term variability for each source also corresponds to the band with the highest count rate.  While we do not present the results of the MOS light curve analysis, we find that the MOS results are in agreement with the pn.

\subsection{Spectral Fitting}
Spectral fitting was done using XSPEC v12.4 \citep{ADASS_96_A}.  For each of the sources, we simultaneously fit the XMM-Newton EPIC (pn + MOS) spectra with the Swift BAT spectra, covering a wavelength range from 0.3--195\,keV.  As an initial fit to the spectra, we used a power law model absorbed by the Galactic column density in the line of sight, in XSPEC {\tt tbabs}*{\tt powerlaw}, to each source (N$_{\rm H} = 1.73 \times 10^{21}$\,cm$^{-2}$ for NVSS 193013+341047 and N$_{\rm H} = 9.83 \times 10^{20}$\,cm$^{-2}$ for IRAS 05218-1212, as obtained from \citealt{1990ARAA..28..215D}).  We added a flux constant to the model to account for differences in the flux between the pn spectrum and each of the MOS and BAT spectra.  This simple model was not a good fit to either spectrum, with $\chi^2/dof =$ 943/330 (reduced $\chi^2 =$ 2.9) for NVSS 193013+341047 and 1184/307 (reduced $\chi^2 = 3.9$) for IRAS 05218-1212.

In \citet{2008ApJ...674..686W} and \citet{2009ApJ...690.1322W}, we describe our choice of a partially covering absorption model ({\tt pcfabs} in XSPEC) to fit more complex spectra in the Swift AGN catalog.  This model, $M(E) = f _{cover}\times \exp^{-N_{\rm H} \sigma(E)} + (1 - f_{cover})$, assumes that the intrinsic absorbing column density only partially covers the AGN emission and that some portion of the direct emission is not blocked by the obscuring region.  Adding this model to each of our source spectra, we find a large statistical improvement in the fit ($\Delta\chi^2$ is 127 for NVSS 193013+341047 and 586 for  IRAS 05218-1212). 

In addition to the partially covering absorption model, we added a reflection model, {\tt pexrav} \citep{1995MNRAS.273..837M}, to account for X-ray emission reflected into the line of sight.  The parameters of this model include the power law index and normalization, which we fixed to the values of the direct emission, the reflection factor ($R = \Omega/2\pi$, where $\Omega$ is the solid angle over which the material is reflected), the source redshift, cutoff energy, and the inclination angle (which we fix at the default of $\cos \theta = 0.45$).  Negative values for the reflection factor indicate a pure reflection spectrum.  Therefore, we allowed this parameter to vary within the range of 0 -- $-5$.  The addition of the reflection model was a significant improvement to the fits for both spectra, with $\Delta\chi^2$ of 24 for NVSS 193013+341047 and 31 for IRAS 05218-1212.  However, we could not constrain the cutoff energy (instead we fixed this to 100\,keV) or the reflection parameters.  

The final component added to our spectral model was a gaussian line ({\tt zgauss}) to fit the Fe K$\alpha$ emission.  A line was statistically significant in both sources, with $\Delta\chi^2$ of 17 for NVSS 193013+341047 and 21 for IRAS 05218-1212.  In addition to the Fe K$\alpha$ emission at 6.41\,keV, the spectrum of IRAS 05218-1212 is more complex.  In Figure~\ref{fig-iron}, we show the ratio of the data/model in the 5--8\,keV band for this source with the best-fit model, excluding the gaussian added to fit Fe K$\alpha$.  The most significant line is clearly the narrow 6.41\,keV emission.  Also present is a narrow emission line at 6.0\,keV.

Full details of our best-fit spectral parameters are presented in Table~\ref{tbl-xray}.  A similar analysis is found in \citet{2011ApJ...736...81T} for IRAS 05218-1212, which uncovers similar best-fit values as our analysis.  To summarize our results from the spectral fits, we find that the spectra of both sources are heavily absorbed with N$_{\rm H} \ge 10^{23}$\,cm$^{-2}$, but Compton thin.  The absorption is partially covering, covering approximately 93\% of the emission in both sources.  The measured power law indices of both sources deviate from the average value of 1.75 found for the Swift BAT sample  \citep{2009ApJ...690.1322W}, with $\Gamma$ being flat for NVSS 193013+341047 (1.42) and steep for IRAS 05218-1212 (2.11).  Reflection is statistically significant in both sets of spectra, but not well constrained.  Additionally, an Fe K$\alpha$ emission line is present in both spectra with an equivalent width of $\approx 150$\,eV.  The width of this line is slightly broad, with $\sigma = 0.14$\,keV, for NVSS 193013+341047.

\section{Comparison with Other Bands}
On the basis of the X-ray spectra alone, these two AGN clearly have complex X-ray spectra.  With higher signal-to-noise X-ray spectra, \citet{2009ApJ...690.1322W} found that more than 50\% of the 153 9-month BAT AGN have complex spectra well-fit by a partial covering model.  Among these, nearly all were absorbed sources ($N_{\rm H} > 10^{22}$\,cm$^{-2}$).  While the initial Swift XRT + BAT fits suggested that these sources were Compton-thick candidates, our XMM-Newton observations reveal that NVSS 193013+341047 and IRAS 05218-1212 are Compton thin.  However, the high column  densities above $10^{23}$\,cm$^{-2}$ and covering fractions $> 90$\% show that these are likely hidden/buried AGN.  This is a new class of objects, comprising 24\% of the Swift AGN, uncovered in high energy surveys, whose AGN nature is hidden in the optical/soft X-rays due to obscuring gas and dust \citep{2007ApJ...664L..79U,2009ApJ...690.1322W,2009ApJ...701.1644W}.  The most interesting properties of these sources, however, are evident when comparing the X-ray properties with other wavelengths.  These include a mis-match between the optical and X-ray classifications and unusually red spectral energy distributions (SEDs).


\subsection{Optical}
Both the optical spectra of NVSS 193013+341047 and IRAS 05218-1212 show clear broad lines, leading to an optical classification of Sy 1--1.8, typically associated with an unabsorbed AGN.  For NVSS 193013+341047, it is unlikely that the X-ray/optical mis-match is due to variability since the broad lines were evident in two spectra taken over 3 months apart (\citealt{2007ApJ...669..109L} and 
\citealt{2010ApJ...710..503W}).  Further, Swift XRT spectra, indicating the X-ray spectrum to be absorbed, were also taken within a few months of the optical.  For IRAS 05218-1212, an EXOSAT spectrum, taken within 2 years of the optical spectrum of \citet{1989ApJ...340..713M}, showed the 2-10\,keV X-ray band to be a factor of 10 times under-luminous compared to other IRAS-selected AGN \citep{1988ApJ...324..767W}, indicating that the source was likely heavily absorbed at this time.  Additionally, an optical spectrum from 22 April 2009, presented in 
\citet{2011ApJ...736...81T}, also confirms the optical broad emission line spectrum.  Thus, the mismatch between X-ray and optical spectra is unlikely due to variability in the sources.  Further, the optical polarization fraction of IRAS 05218-1212 was reported as less than 1\% \citep{1990MNRAS.244..604B}, showing that the optical light is not mostly scattered around the X-ray obscuring region.

In the XMM large scale structure survey, 7 out of 61 (11\%) optical broad line sources exhibited X-ray absorbed spectra \citep{2007AA...474..473G}.  It is interesting to note that all of these sources are more luminous (in L$_{2 - 10 keV}$ by $\approx 100$) and at much higher redshift ($<z> = 1.93$) than our target sources.  Similar sources have been found by \citet{2004AA...421..491P} and 
\citet{2006ApJS..165...19E}, comprising 5 -- 15\% of their X-ray samples.  Thus, our targets may be the first low-redshift AGN discovered from this class of object, which includes a significant fraction of high redshift AGN.  As the low-redshift analogs of this class, we will be able to study the sources and hosts, in particular, in much greater detail due to their much closer distance.

\subsection{Infrared}
Not only are the X-ray and optical spectra of our targets puzzling, but both sources are very red, with $J-K =$ 2.43 (NVSS 193013+341047) and 1.74 (IRAS 05218-1212).  These values are high compared to the 9-month BAT AGN sample, taken from NED.  In Figure~\ref{fig-jk}, we plot the 2MASS $J-K$ values for the 9-month sample versus both X-ray absorbing column and the 14--195\,keV luminosity, obtained from \citet{2009ApJ...690.1322W}.  These plots show a few interesting results.  First, Sy1s (triangles) are redder than Sy2s (circles), with the average $J-K$ of 1.32 for the 31 Sy 1--1.5s, 1.13 for the 26 Sy 1.5--1.9s, and 1.07 for the 30 Sy2s.  The high $J-K$ values of NVSS 193013+341047 and IRAS 05218-1212 align them more closely with the Sy1s.  This also causes our sources, along with 3C 105, to inhabit a strange place in the plot of $J-K$ vs. N$_{\rm H}$, being higher column density sources with higher $J-K$ values.  3C 105, however,  is unlike our sources in that it is a narrow line radio galaxy which does not exhibit broad optical lines.

Another interesting result we find is that the sources with the highest $J-K$ values are also the most luminous of the BAT sample.  In addition to NVSS 193013+341047 and IRAS 05218-1212,  3C 105, 3C 111.0, 3C 273, and EXO 055620-3820 are the most luminous sources with the highest $J-K$.  Among these, only 3C 105 has a high column density.  The host galaxies of both 3C 105 and 3C 111.0 are identified as ellipticals from HST observations with NICMOS \citep{2006ApJS..164..307M}.  3C 273 is a blazar, while EXO 055620-3820 is a Sy1 with a complex X-ray spectrum best-fit by an ionized partial covering model \citep{1996ApJ...463..134T}.  The host types are unknown for both of these sources.
Both NVSS 193013+341047 and IRAS 05218-1212 are unusual, even among the most luminous and highest $J-K$ sources in the sample, in having a mismatch between the X-ray and optical types.  While the X-ray spectra are similar to 3C 105, this source exhibits no optical broad lines.

\subsection{Broadband SEDs}
In Figure~\ref{fig-sed}, we show the observed broad-band SEDs for both NVSS 193013+341047 and IRAS 05218-1212.  The radio/IR data includes publicly available flux densities from NVSS, IRAS, and 2MASS (obtained through the NASA/IPAC Extragalactic Database).  In addition to this, we include our own measurements from Swift BAT and XMM-Newton -- including both the observed X-ray measurements based on our best-fit model and Optical/UV measurements from the Optical Montior (OM) and the Swift UV/optical telescope (UVOT).  

Data from the OM were available in the U and UVW1 filters for IRAS 02218-1212 and in the UVW1 filter for NVSS 193013+341047.  We processed the OM data with the SAS script {\tt omichain}.  The flux and magnitudes were extracted with the interactive SAS package {\tt omsource}, using a 6\arcsec\,circular aperture centered on the source and a circular background region, free of sources.   We also include Swift UVOT observations for our target sources, which were obtained using the same region files as for the OM data, and extracted with the Swift tool {\tt uvotsource}.  Processed Swift UVOT observations were available in the V, B, U, UVW1, UVM2, and UVW2 bands for NVSS 193013+341047 (with Swift observations 00035274002 and 00035274003) and the V, UVM2, and UVW2 bands for IRAS 02218-1212 (Swift observations 00037085001, 00037085002, 00037085003, and 00037085004).  Our broadband SEDs also include the X-ray flux densities measured in the 0.5--2\,keV, 2--10\,keV, and 14--195\,keV bands, based on our joint XMM + BAT spectral fits.  The measurements are included in Table~\ref{tbl-sed}.

\begin{sloppypar}
The spectral energy distributions (SEDs) for both sources, particularly for NVSS 193013+341047, show the sources to be very bright in the IR with sharp declines in flux into the UV and soft X-rays.  While bright IR emission could be the result of reprocessing of the AGN emission in dust, which obscures the optical/soft X-rays, it is still unclear as to why there are optical broad lines.
We are unaware of any other low-redshift AGN of type 1 or 2 with similar SEDs.  While other red AGN ($J - K > 2$) have been selected through 2MASS, X-ray follow-ups have not shown the same mismatch between X-ray and optical spectra \citep{2005ApJ...634..183W}.
\end{sloppypar}

\section{Discussion}


\subsection{Optical and X-ray Reddening}
Multi-wavelength observations of active galaxies previously revealed differences in the reddening measured between X-ray and optical spectroscopy.  Assuming a standard Milky Way extinction curve and dust-to-gas ratio, derived X-ray column densities can be up to ten to one hundred times that of estimates based on optical emission lines (e.g., \citealt{1982ApJ...256...92M,1982ApJ...257...47M,2001AA...365...28M}).  In the Swift sample, optical estimates of the reddening from the ratio of narrow H$\alpha$/H$\beta$ emission lines show this same trend with the X-ray derived hydrogen columns 10--100 times larger than the optically derived column densities \citep{2010AIPC.1248..369W}.  For NVSS 193013+341047 in particular, our estimate of the reddening from the ratio of narrow H$\alpha$/H$\beta$ is $E(B-V) = 0.16$ \citep{2010ApJ...710..503W}, corresponding to an optically derived column density of $8.5 \times 10^{20}$\,cm$^{-2}$ (based on the conversion from \citealt{2001AA...365...28M}).  Therefore, for this source the X-ray derived column is nearly 300 times larger than the optically derived column, on the extreme end of what is typically found for differences in the optical/X-ray derived reddening values.

There are a number of ideas in the literature about how to account for differences in the X-ray/optical obscuration.  One possible explanation for mismatches in the optical/X-ray reddening is that the dust-to-gas ratio for AGN is lower than in the ISM in our own Galaxy.  \citet{2001A&A...365...37M} suggest that large dust grains in the region around the AGN are responsible for the differences between the estimates of the optical and X-ray reddening.  Among the evidence for unusual dust properties, \citet{2001AA...365...28M} show that active galaxies have faint to no detection of the silicate features at 9.7\,$\mu$m and the 2175\,\AA~carbon feature, both of which indicate the presence of small dust grains.  Several other possibilities to explain the optical/X-ray mismatch that are discussed in these papers include higher metallicities in active galaxies, the hard X-rays being absorbed by the optical broad emission line producing clouds (which are expected to have N$_{\rm H} \sim 10^{23}$\,cm$^{-2}$), and scattering of the broad line emission through a region of lower column density than what obscures the X-ray emission.  Alternatively, \citet{2006ApJ...653..127S} propose an accretion disk geometry that can explain the X-ray obscuration and 9.7\,$\mu$m silicate feature.  Yet another explanation for differences in reddening is that the X-ray absorption is related to an AGN-driven wind.  We describe this possibility in the following sub-section.

\subsection{Ionized Emitters}

Mis-matches in classification between the X-ray absorbed and broad-line optical spectra are sometimes explained by the presence of a high column of ionized gas.  Such a model has been adopted to account for active galaxies at both low and high redshift.  For instance, the optical and UV spectra of the low-z AGN NGC 4151 reveal broad emission lines while the X-ray grating observations show that the soft X-rays are heavily obscured and dominated by strong ionized emission (e.g.,  \citealt{2005ApJ...633..693K}).  NGC 4151 is classified similarly as a Sy 1.5, with a strongly detected multiple-component ionized outflowing wind in the UV/X-rays  \citep{2005ApJ...633..693K,2006ApJS..167..161K}.  NGC 4151 is also detected in the Swift catalog, however, both the amount of reddening (from J-K) and the column density of gas are much lower than the two new sources we present in this current paper, showing that it is not a useful analog for heavily reddened NVSS 193013+341047 and IRAS 05218-1212.  

Perhaps a more relevant example of a source with a mis-match between optical and X-ray spectra is the broad-line radio galaxy 3C 445, which is at a similar redshift to our sources with $z = 0.057$.  Broad emission lines are observed in this source (FWHM $\sim 6400$\,km\,s$^{-1}$) along with a highly reddened continuum (E$_{\rm B - V} = 1$\,mag; \citealt{1988AJ.....96.1208C}).  Suzaku and Chandra observations show that 3C 445 is heavily absorbed with a column density of $\sim 10^{23}$\,cm$^{-2}$ and that the soft spectrum, like that of NGC 4151, is dominated by emission lines \citep{2011MNRAS.414.2739B}.  \citet{2011MNRAS.414.2739B} conclude that the photo-ionized emitting gas is likely associated with the BLR gas, which we are able to view because the high column density absorber is not completely blocking the BLR.  This is possible if the obscuring region is not a uniform torus but instead clumpy or wind-like in structure (e.g., \citealt{2002ApJ...571..234R} and \citealt{2008NewAR..52..274E}).

In both NVSS 193013+341047 and IRAS 05218-1212 the residuals in the soft X-rays of our spectra suggest that emission lines may be present, but our observations do not have the resolution to confirm them.  If higher resolution X-ray observations indicate both the high column density neutral obscuration and a strong emission component, then the same clumpy torus/wind models used to explain NGC 4151 and 3C 445 also explain the mis-match in classifications for our unusual sources.  To confirm strong emission, potentially associated with an outflow, we will propose for grating observations of these sources with XMM-Newton/Chandra.

Recently, active galaxy/quasar driven outflows were also invoked to explain highly absorbed QSOs detected with ROSAT/sub-mm surveys 
\citep{2011MNRAS.416.2792P}.  In this case, the $z \sim 2$ QSOs with rest-frame broad UV emission lines and heavily absorbed X-rays have X-ray spectra that can be modeled with ionized absorption.  \citet{2005ApJ...632..736A} suggest that these sources may be in a transition between the absorbed and unabsorbed phases.  As \citet{2011MNRAS.416.2792P} show, this stage may be marked by strong AGN-driven winds.  If this is the case, then the intermediate Seyfert 1 types, like the newly detected Swift sources we discuss in this paper, are likely in a similar transitional stage.  This is supported by the fact that many of the well-known Seyferts with strong outflow detections are Sy 1.5s (e.g., NGC 3516, NGC 3783, and NGC 4151).  

\subsection{Low-redshift EROs?}
While there are no known low-redshift AGN with similar SEDs to NVSS 193013+341047 and IRAS 05218-1212, the high redshift universe does provide a useful analog.  Namely, we find that the SEDs of our sources, particularly NVSS 193013+341047, match that of XBS J0216-0435 ($z \approx 2$).  This source is identified as a type 2 quasar hosted in an ERO, whose SED is presented in \citet{2006AA...451..859S}.  In Figure~\ref{fig-sed}, we include the SED of this source with a cross symbol.  As pointed out and illustrated in Figure 6 of \citet{2006AA...451..859S}, the SED is unlike those of radio quiet or radio loud AGN.

EROs are identified as red sources with $R-K \ga 5$ \citep{1988ApJ...331L..77E}
or $I-K \ga 4$.  Joint IR/X-ray studies of EROs have found that most AGN in this category are heavily obscured ($N_{\rm H} > 10^{22}$\,cm$^{-2}$) \citep{2002AJ....123.1149A}.  Based on the 2MASS (J, H, K) colors, we estimate $R-K$ of 4.8 (NVSS 193013+341047) and 3.5 (IRAS 05218-1212).  This places them below the formal definition of an ERO.  However, with a similar SED, very red colors, and X-ray absorbed spectra, our sources may be the low redshift equivalent.  Particularly, the result showing $J-K$ values to increase with increasing luminosity (Figure~\ref{fig-jk}), supports this claim, as our sources are a factor of  10--100 less luminous in the 2--10\,keV X-ray band.  If these sources are low redshift EROs, then multi-wavelength studies of their properties will help to answer many of the questions surrounding these high redshift sources. 

Previous X-ray observations of EROs suggest that they are absorbed sources with N$_{\rm H} > 10^{22}$\,cm$^{-2}$, representing a class of type II quasars \citep{2004A&A...418..827M,2005A&A...432...69B}.  These type II quasars all tend to be hosted within elliptical galaxies.  While we can not be sure of the morphology of NVSS 193013+341047 and IRAS 05218-1212, no spiral arms are distinguished in the available Digital Sky Survey images.  The morphologies of Swift-detected AGN with similar $J-K$ colors all tend to be in ellipticals, so it is possible that our unusual sources are also hosted in ellipticals.  If so, our highly reddened active galaxies are two of only a handful of Swift-detected AGN with elliptical hosts, as the majority of the very hard X-ray detected galaxies are associated with spirals \citep{2011ApJ...739...57K}.

\section{Summary}
In this paper, we present two very unusual AGN discovered by the Swift BAT all-sky survey.  In the X-ray band, these sources are clearly heavily absorbed (N$_{\rm H} =$1--2 $\times 10^{23}$\,cm$^{-2}$), with little X-ray flux below 2\,keV.  Both of these sources exhibit broad lines in their optical spectra, seemingly at odds with the heavily absorbed X-ray spectra.  Further, the SEDs, revealed through publicly available NVSS, IRAS, 2MASS, and Swift data, are very red.  As such, these sources appear to be the low-redshift analogs of AGN hosted in EROs.   It is still unclear, however, how the multi-wavelength properties fit together, particularly the mismatch between X-ray and optical classification, but one possible explanation is that the heavy obscuring column density is only partially obscuring the BLR, as well as an associated X-ray emitting region associated with an outflow.  High signal-to-noise grating observations are required to confirm outflowing ionized gas in the spectra of these sources.

Another possible explanation for the mis-match in classifications is that the dust properties in the host galaxies of NVSS 193013+341047 and IRAS 05218-1212 are different.  Infrared spectroscopy could help to uncover the dust properties in these sources.  What we can discern, on a larger scale, is that the host galaxy types of these unusual sources may be different from the majority of the BAT-selected sample.  From the 2MASS derived absolute magnitudes of $J = -22.8$ and $-23.78$, we find that the host galaxies are of typical luminosity (L$_*$).  Unfortunately, the available images from UVOT and the Digital Sky Survey, are not of sufficient quality to resolve structure, determine whether the sources are actively undergoing star formation, or discern dust lanes.  Since spiral arms are not indicated and two of the BAT AGN with high 14--195\,keV luminosities and similar $J-K$ values are hosted in ellipticals, it is possible that the hosts of these sources are also ellipticals.  If this is the case, the sources are particularly interesting, being among only $\approx 5/100$ BAT AGNs known to be hosted in an elliptical.

Finally, and possibly most importantly, we showed that the 2MASS-derived $J-K$ values for the BAT AGN appear to be connected to both the amount of X-ray obscuration and the 14--195\,keV luminosity.  We showed that $J-K$ is lower for Sy2s and higher for Sy1s as well as for high luminosity sources.  This is interesting considering that at high redshift ($z \approx$ 1--2) the reddest EROs are now being used as a means to select obscured Type II QSOs (e.g., \citealt{2005AA...432...69B}).

\acknowledgments
The authors thank the anonymous referee for useful comments that improved our discussion.  We gratefully acknowledge support for this work through NASA XMM-Newton guest observer grant  NNX09AP79G.  LMW also acknowledges support through NASA grant HST-HF-51263.01-A, through a Hubble Fellowship from the Space Telescope Science Institute, which is operated by the Association of Universities for Research in Astronomy, Incorporated, under NASA contract NAS5-26555.
This work utilizes observations obtained with XMM-Newton, an ESA science mission with instruments and contributions directly funded by ESA Member States and NASA.  This research has made use of data obtained through the High Energy Astrophysics Science Archive Research Center Online Service, provided by the NASA/Goddard Space Flight Center.  Additionally, this research has made use of the NASA/IPAC Extragalactic Database (NED) which is operated by the Jet Propulsion Laboratory, California Institute of Technology, under contract with the National Aeronautics and Space Administration.

\bibliography{/Users/lisa/Documents/Documents/MyBibtex}

\begin{deluxetable}{clllllcc}
\tabletypesize{\scriptsize}
\tablecaption{Details of the XMM-Newton Observations\label{tbl-1}}
\tablewidth{0pt}
\tablehead{
\colhead{Source} & \colhead{RA} & \colhead{Dec} & \colhead{z} & \colhead{Obs. ID} & \colhead{Date} & \colhead{Exp. Time\tablenotemark{\dagger}} &\colhead{$<$Ct Rate$>$\tablenotemark{\dagger}}
}
\startdata
\object{NVSS 193013+341047} & 19:30:13.86 & +34:10:50.04 & 0.0629 & 0602840101	 & 16-05-2009 & 12.91 & 0.29\\
\\
\object{IRAS 05218-1212} & 05:24:06.80 & -12:10:10.87 & 0.0490 & 0551950401 & 24-08-2008 & 11.27 & 0.17 \\
\enddata 
\tablenotetext{\dagger}{The exposure time (in units of ks) and count rate (in units of cts\,s$^{-1}$) are recorded for the pn spectra in the 0.1--10\,keV band.}
\end{deluxetable}

\begin{deluxetable}{ccc}
\tabletypesize{\small}
\tablecaption{Joint XMM-Newton and BAT spectral fits\label{tbl-xray}}
\tablewidth{0pt}
\tablehead{
\colhead{Parameter} & \colhead{NVSS 193013+341047} & \colhead{IRAS 05218-1212}
}
\startdata
$N_{\rm H}$\tablenotemark{a} & $24.5^{+3.7}_{-4.8}$ & $10.9^{+0.81}_{-0.77}$\\
$f_{cover}$\tablenotemark{a}  & $0.93^{+0.02}_{-0.03}$ & $0.93^{+0.01}_{-0.01}$ \\
$\Gamma$\tablenotemark{b} & $1.42^{+0.10}_{-0.13}$ & $2.11^{+0.09}_{-0.10}$ \\
$A_{\Gamma}$\tablenotemark{b} & 9.85 $\times 10^{-4}$& 1.74 $\times 10^{-3}$\\
R\tablenotemark{b} & (-5) & (-5)\\

E$_{Fe K\alpha}$\tablenotemark{c} & 6.38$^{+0.07}_{-0.08}$& 6.41$^{+0.06}_{-0.06}$\\
$\sigma$\tablenotemark{c}  & 0.14$^{+0.13}_{-0.08}$ & 0.07$^{+0.06}_{-0.07}$\\
EW\tablenotemark{c}  & $149^{+79}_{-65}$ & $172^{+137}_{-63}$ \\

$\chi^2/dof$ & 292/323 & 302/300\\
\enddata
\tablenotetext{a}{Parameters of the {\tt pcfabs} model, including hydrogen column density measured in units of $10^{22}$\,atoms\,cm$^{-2}$ and the covering fraction of the absorber.}
\tablenotetext{b}{Spectral index computed with the reflection model {\tt pexrav} assuming total reflection and a fixed energy cutoff at 150\,keV.  The reflection parameter is not well-constrained.  The quoted normalization on the power-law component is in units of photons\,keV$^{-1}$\,cm$^{-2}$\,s$^{-1}$ at 1\,keV.}
\tablenotetext{c}{Equivalent width and normalization (with errors or as an upper limit) on a fixed Gaussian Fe K line at 6.41\,keV in units of eV and $10^{-5} \times$ total photons\,cm$^{-2}$\,s$^{-1}$, respectively.}
\
\end{deluxetable}

\begin{deluxetable}{lc | cc}
\tabletypesize{\small}
\tablecaption{Multi-wavelength Flux Measurements\label{tbl-sed}}
\tablewidth{0pt}
\tablehead{
\colhead{SED Data} & \colhead{} & \colhead{NVSS 193013+341047} & \colhead{IRAS 05218-1212} \\
\hline \\
\colhead{Measurement} & \colhead{$\nu$ (Hz)} & \colhead{F$_{\nu}$ (mJy)} & \colhead{F$_{\nu}$ (Jy)} \\
}
\startdata
NVSS & 1.4$\times 10^9$ & 4.4 $\pm 0.6$ & 9.2 $\pm$ 1.1 \\
IRAS 100$\mu$  & 3$\times 10^{12}$&  \nodata & 1020 \\
IRAS 60$\mu$  & 5$\times 10^{12}$ & \nodata & 456.3 $\pm$ 50 \\
IRAS 25$\mu$   & 1.2$\times 10^{13}$&  \nodata & 280 \\
IRAS 12$\mu$ &  2.5$\times 10^{13}$ & \nodata & 100 \\
2MASS K$_s$ & 1.388571$\times 10^{14}$ & 12.6 & 16.5 \\
2MASS H &  1.803805$\times 10^{14}$ & 6.05 & 11.3 \\
2MASS J & 2.427469$\times 10^{14}$ & 3.2 & 7.95 \\
UVOT B & 6.662055$\times 10^{14}$ & 0.277 $\pm 0.049$ & \nodata \\
UVOT B & 6.662055$\times 10^{14}$ & 0.26 $\pm 0.01$ & \nodata \\
OM U & 8.30781414$\times 10^{14}$ & \nodata & 2.41 $\pm 0.10$ \\
UVOT V & 8.446$\times 10^{14}$ & 0.78 $\pm0.09 $ & 6.15 $\pm 0.10$\\
UVOT U & 8.81742524$\times 10^{14}$ & 0.13 $\pm 0.02$ & \nodata \\
UVOT U & 8.81742524$\times 10^{14}$ & 0.196 $\pm 0.019$ \\
OM UVW1 & 9.82092118$\times 10^{14}$ & 0.05 $\pm 0.01$ & 2.10$\pm 0.01$ \\
UVOT UVM2 & 1.297803$\times 10^{15}$ & 0.027 $\pm 0.006$ & 1.49 $\pm 0.01$ \\
UVOT UVM2 & 1.297803$\times 10^{15}$ & 0.024 $\pm 0.007$ & 1.64 $\pm 0.02$ \\
UVOT UVW2 & 1.414115$\times 10^{15}$ & 0.030 $\pm 0.005$ & 1.41 $\pm 0.02$ \\
UVOT UVW2 & 1.414115$\times 10^{15}$ & \nodata & 1.54 $\pm 0.02$ \\
PN soft & 1.25643545$\times 10^{17}$ & 0.00010 & 0.00036 \\
PN hard & 1.89640880$\times 10^{18}$ & 0.00015 & 0.00027 \\ 
BAT & 6.01787337$\times 10^{18}$ & 0.00013 &  0.00029 \\
\enddata
\vspace{0.1cm}
Notes: The NVSS, IRAS, and 2MASS flux densities were obtained using the NASA/IPAC Extragalactic Database.  Swift UVOT and XMM-Newton OM observations were analyzed as explained in the text.  The X-ray measurements, obtained from the best-fit X-ray models, correspond to the 0.5-2 keV soft band, 2-10 keV hard band, and the 14-195 keV BAT band fluxes.
\end{deluxetable}

\begin{figure}
\centering
\includegraphics[width=12cm]{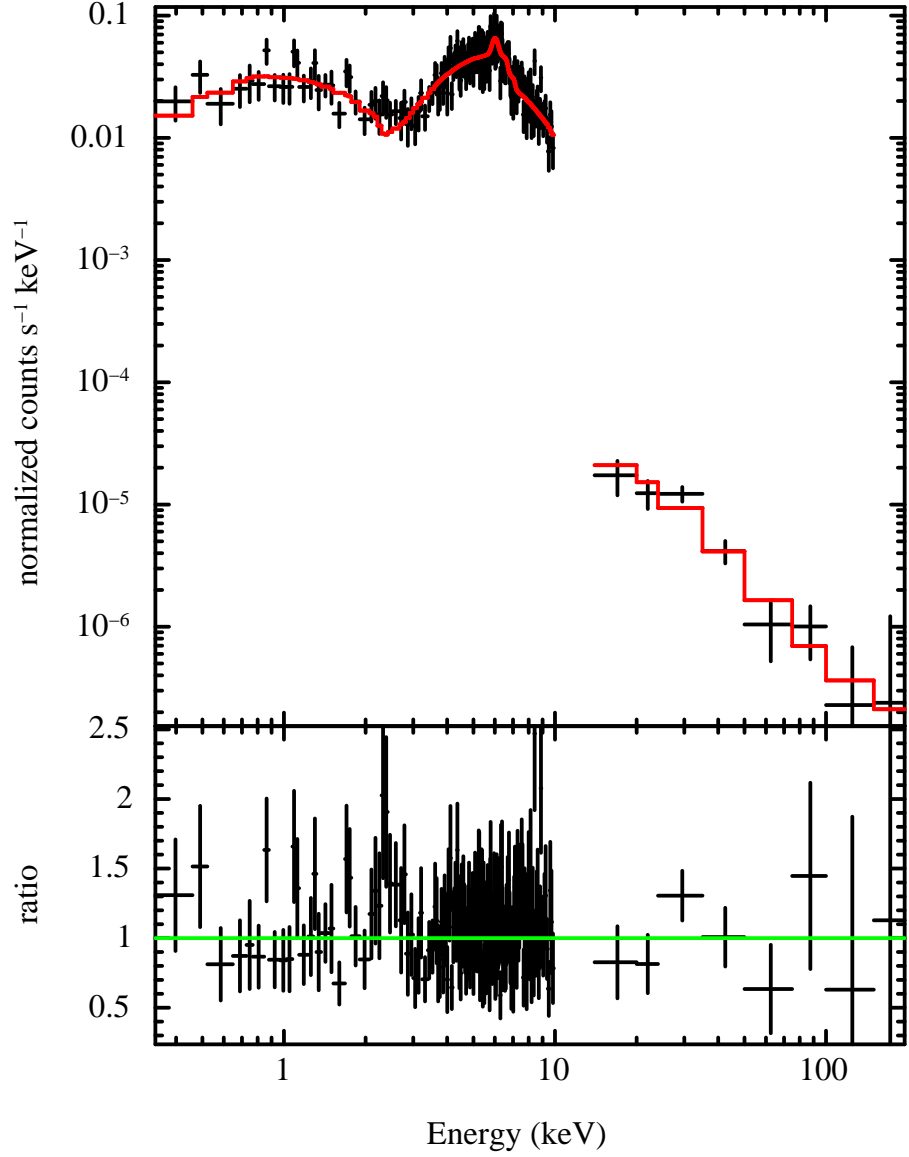}
\caption{The XMM-Newton pn and Swift BAT spectra are plotted for NVSS 193013+341047, along with the best fit model, as well as the ratio of the data/model in the bottom plot.  The spectrum is complex and required a partial covering absorber model.
\label{fig-nvss_spec}}
\end{figure}

\begin{figure}
\centering
\includegraphics[width=12cm]{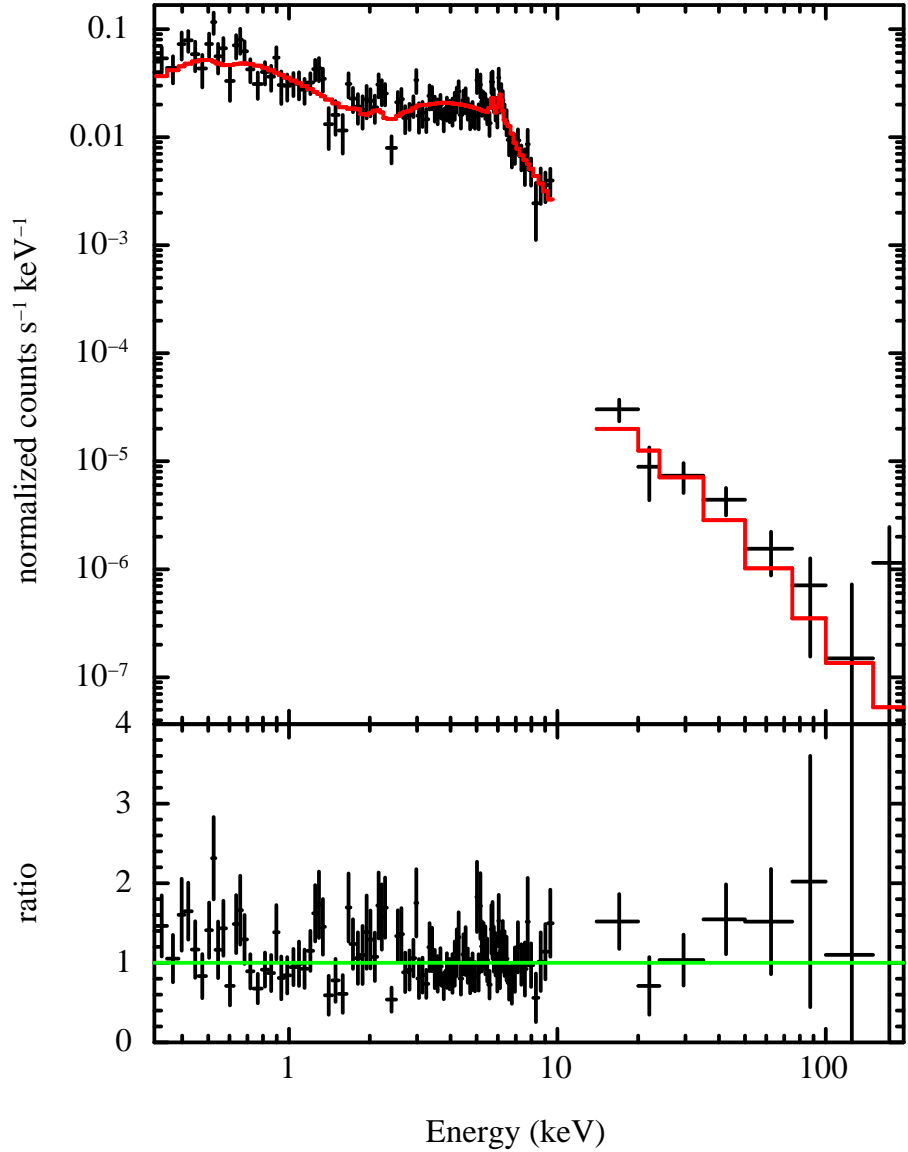}
\caption{The XMM-Newton pn and Swift BAT spectra are plotted for IRAS 05218-1212, along with the best fit model, as well as the ratio of the data/model in the bottom plot.  The spectrum is complex and required a partial covering absorber model.
\label{fig-iras_spec}}
\end{figure}

\begin{figure}
\centering
\includegraphics[width=12cm]{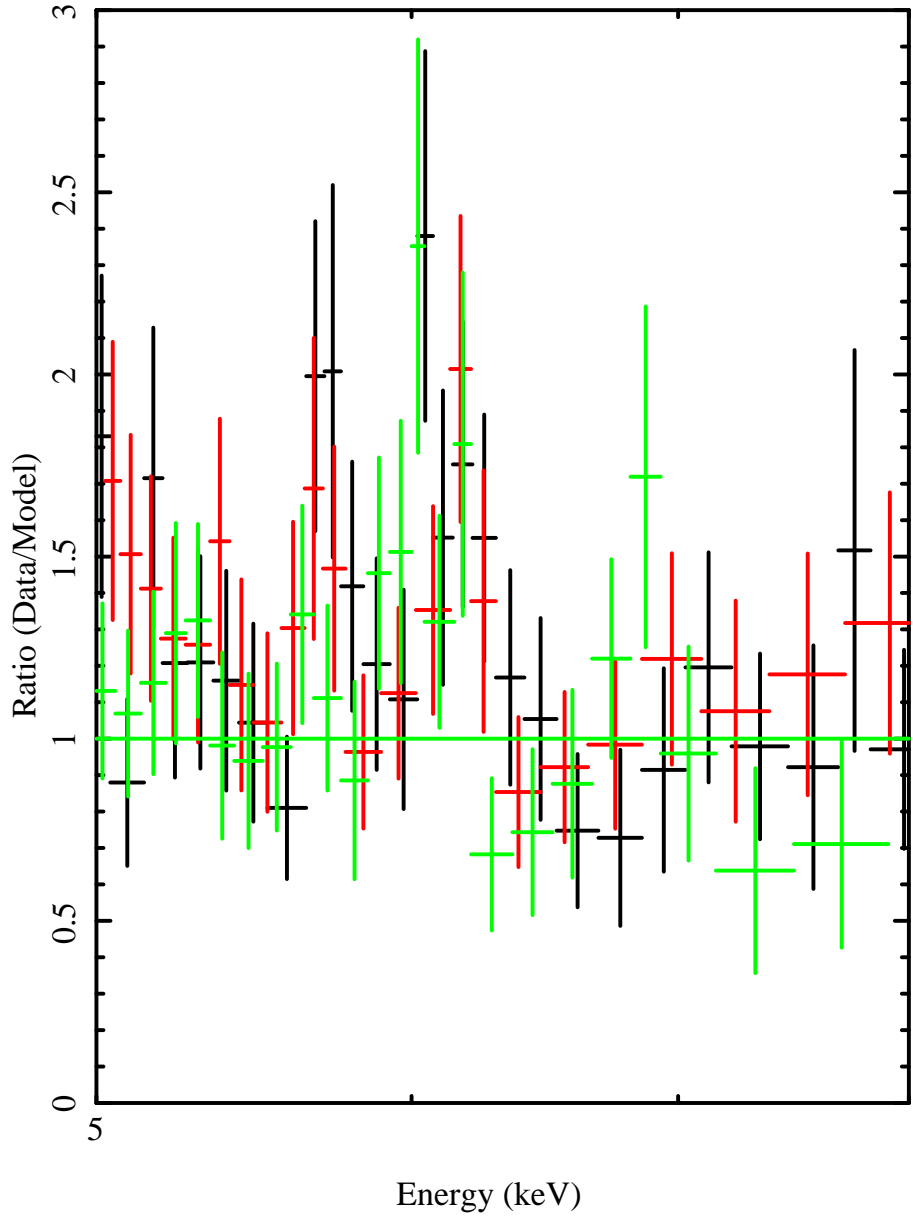}
\caption{The ratio of the data/model for a partially covering power law model without accounting for the Fe K$\alpha$ emission is shown in the 5--8\,keV band.  The pn (black), MOS1 (red), and MOS2 (green) data points are shown.  In addition to the narrow Fe K$\alpha$ line at 6.41\,keV, a second line is significant in each of the detectors ($\Delta\chi^2 = 9$) at 6.0\,keV.
\label{fig-iron}}
\end{figure}

\begin{figure}[hbt]
\centering
\includegraphics[width=5in]{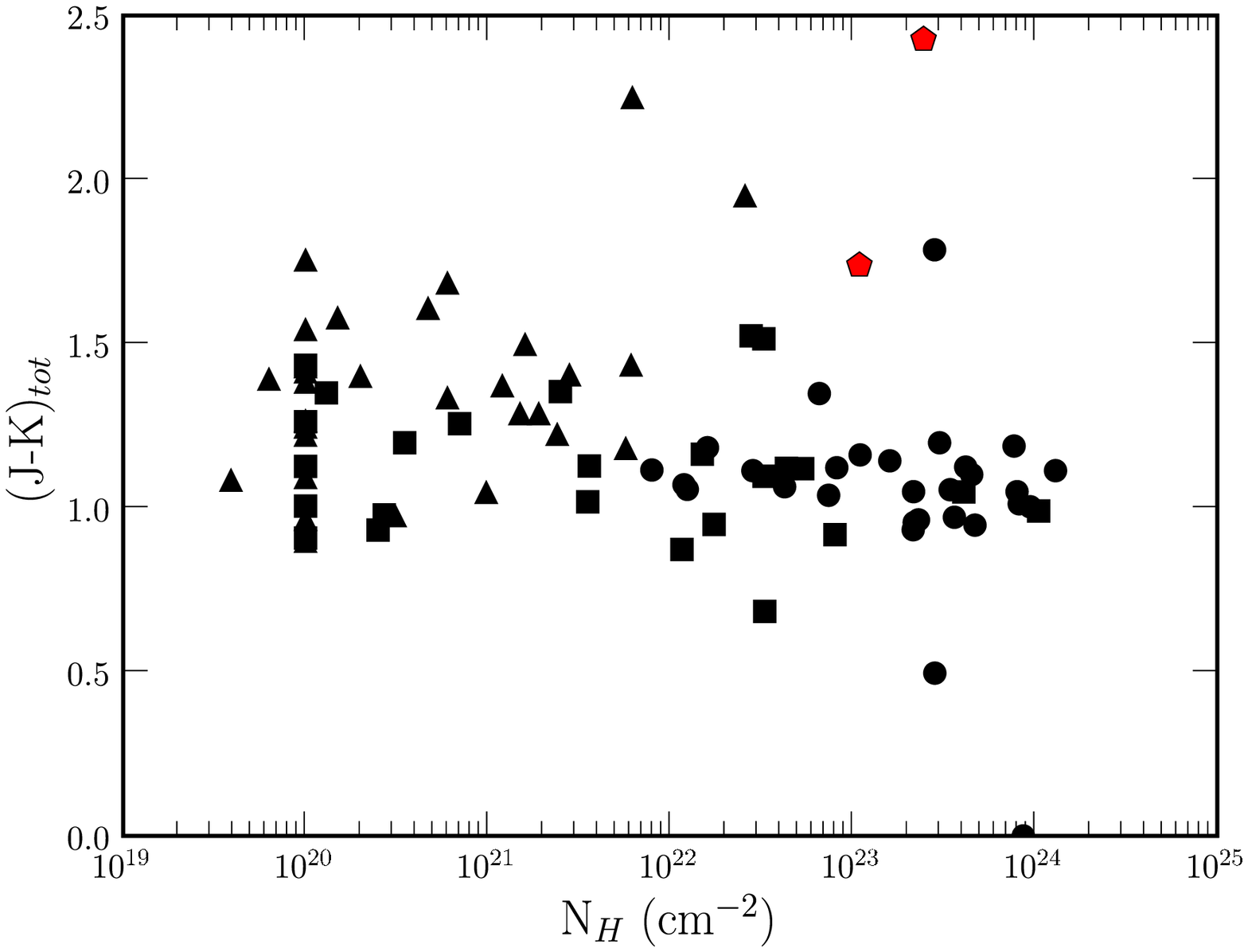}
\includegraphics[width=5in]{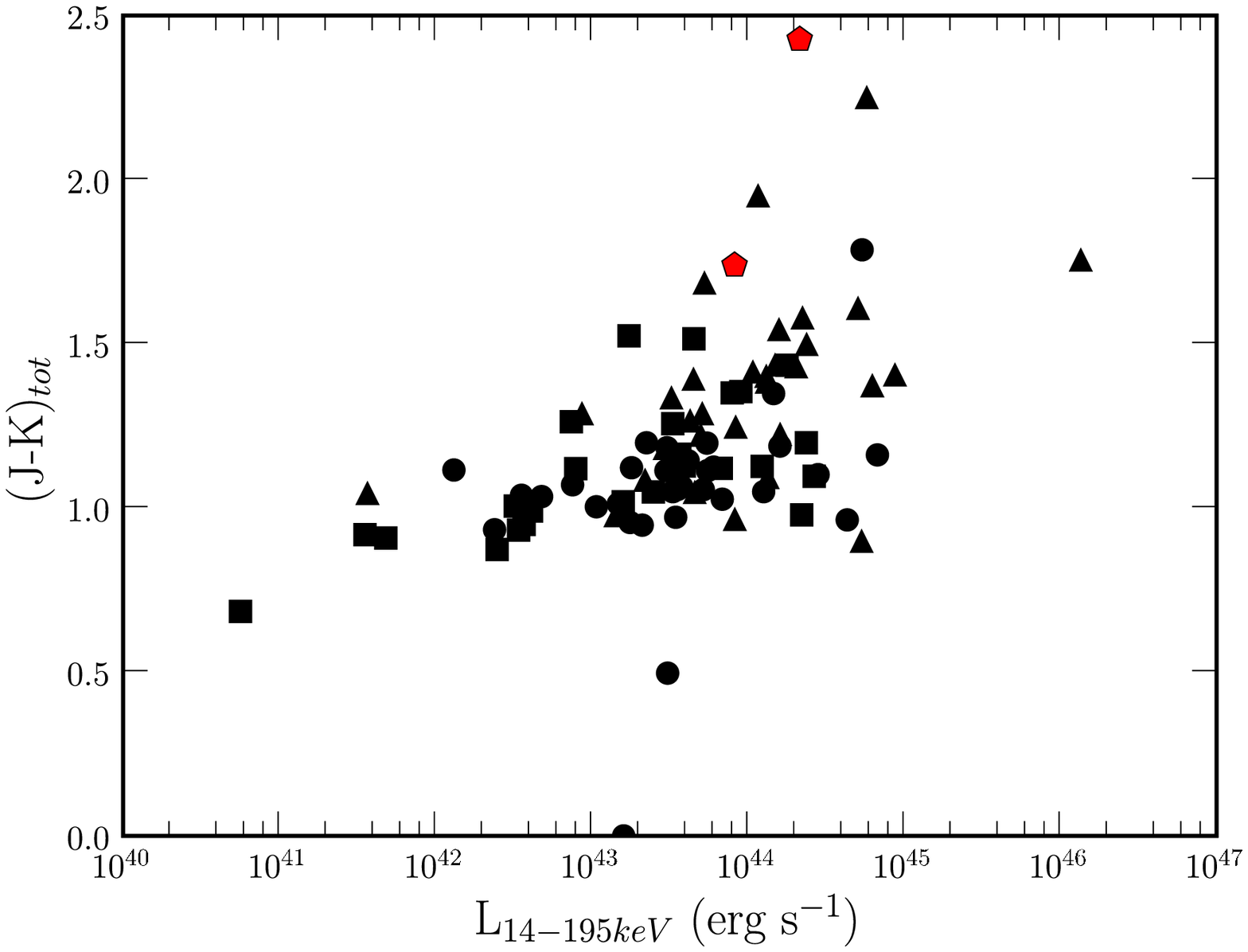}
\caption{Plot of reddenning versus the X-ray measured column density (N$_{\rm H}$) and the 14--195\,keV luminosity from BAT.  The reddenning (J-K)
is constructed from the 2MASS values, publicly available in NED.   NVSS 193013+341047 and IRAS 05218-1212 (pentagons) are among the reddest sources, along with 3C 111.0, EXO 055620-3820, 3C 105, and 3C 273.  Seyfert 1--1.5 sources (triangles), Seyfert 1.5--2 (squares), and Seyfert 2s (circles) are all indicated.\label{fig-jk}}
\end{figure}

\begin{figure}[hbt]
\centering
\includegraphics[width=6in]{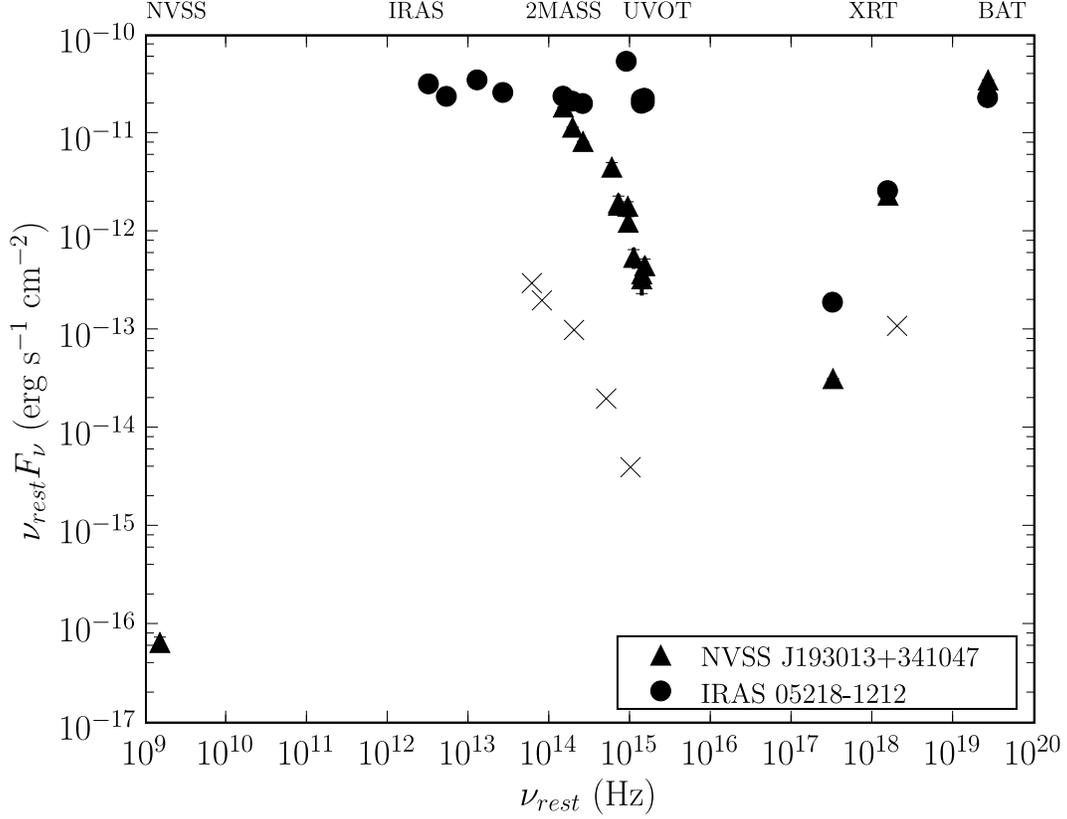}
\caption{We constructed the observed spectral energy distributions (SEDs) for our targets using available radio/IRAS/2MASS measurements as well as our own measurements from the SWIFT and XMM-Newton observatories (including optical/UV from UVOT and XMM OM, 0.5 -- 10\,keV X-rays with XRT/XMM pn, and 14 -- 195\,keV very hard X-rays with BAT).  Both sources show a steep decline in flux from infrared into ultraviolet and soft X-rays, not typical of the other BAT AGNs.  However, the sharp drop in flux from the IR to UV/soft X-rays is similar to the SED of ERO source XBS J0216-0435 \citep{2006AA...451..859S}, plotted with the cross symbols.\label{fig-sed}}
\end{figure}

\end{document}